# Documenting Problem-Solving Knowledge: Proposed Annotation Design Guidelines and their Application to Spreadsheet Tools

Matthew Dinmore
Department of Information Systems, University of Maryland, Baltimore County
1000 Hilltop Circle, Baltimore, MD 21250
mdinmo1@umbc.edu

**ABSTRACT**

*End-user programmers create software to solve problems, yet the problem-solving knowledge generated in the process often remains tacit within the software artifact. One approach to exposing this knowledge is to enable the end-user to annotate the artifact as they create and use it. A 3-level model of annotation is presented and guidelines are proposed for the design of end-user programming environments supporting the explicit and literate annotation levels. These guidelines are then applied to the spreadsheet end-user programming paradigm.*

## 1. INTRODUCTION

End-user programmers' principal goal is to solve problems, not write programs. However, the mechanics of solving problems through programming – even when the user is not intentionally writing code, as in traditional programming – often take precedence over documenting the solution. The resulting software artifact contains within it the tacit problem-solving knowledge gained in the process of not only creating it, but of putting it into practical use. Because this knowledge is rarely made fully explicit, reusability as both a software and knowledge artifact is reduced.

A significant challenge is finding a means to document problem-solving knowledge in an unobtrusive, seamless manner. If the cost of capturing the knowledge is greater than the perceived benefit, a user is unlikely to do it. This is largely a matter of design; regular users of a tool will know the challenge of deciphering an artifact they themselves created, so there is certainly incentive to document for themselves. Those engaged in collaborative problem-solving endeavors have an incentive to share, and a design that enables, at a minimum, the kind of simple, in-process note-taking common to this work [Markus, 2001] is a worthy objective toward satisfying this requirement.

An apt place to start is with the spreadsheet. It is by far the most popular end-user programming tool [Jones et al., 2003]; its widespread use in the financial industry and other domains has led to the recognition that errors can be costly [Panko, 1998]; these are certainly software errors, and as such, there are efforts to bring software engineering-like practices to the creation of spreadsheets and other end-user development media [Burnett et al., 2004].

A common solution that potentially satisfies both requirements – to capture knowledge and reduce errors – may be found in designs supporting structured annotations. Knowledge is made explicit through annotation [Marshall, 1998] and there is evidence that structured combinations of code and documentation reduce errors and improve maintainability [Oman and Cook, 1990]. The objective of this work is to combine and extend these ideas in the end-user development space, with an initial focus on the spreadsheet paradigm.

This paper begins by scoping the work around problem-solving knowledge transfer and typical methods of achieving this in software development. A model of end-user development annotation is introduced as a framework for a collection of guidelines for integrating annotation as a means of knowledge documentation into the end-user programming paradigm. These guidelines are then applied to the spreadsheet end-user development modality, and a plan for evaluating the resulting designs is described. Finally, related work is compared to the proposed approach.

## 2. PROBLEM-SOLVING KNOWLEDGE

This work focuses on improving the transfer of problem-solving knowledge. This is formally defined as the ability to apply previously learned knowledge to solving a new problem [Mayer and Wittrock, 1996]. In this process, prior learning helps new learning; this effect is increased when the prior learning – that is, the hard-won problem-solving knowledge learned by another – is codified in such a way to make it useful to someone taking on a new problem. A common approach to this is analogical problem solving, in which the solution to a previous, but analogous, problem is tailored to the new problem. The challenge then becomes in finding analogous problems and solutions to reuse.

The degree of codification of the knowledge is closely related to the concepts of tacit, implicit and explicit knowledge [Polanyi, 1966; Wilson, 2002]. Briefly, tacit knowledge is that which is unexpressed and largely inexpressible; implicit knowledge is partially expressed but may require additional knowledge – some "common sense," at least as far as the expresser of the knowledge is concerned – to understand it; and explicit knowledge is fully stated and generally reusable as is. Many challenges face efforts to effectively represent and share knowledge, and these are often beyond the purely technical, requiring engineering in the larger socio-technical domain [Fischer and Otswald, 2001].

To scope this challenge, a useful point of departure is to identify the elements of problem-solving knowledge that need to be communicated. As adapted to computer-based problem solving by Mayer [2002], there are four kinds of knowledge of interest to knowledge transfer: declarative, conceptual, procedural and metacognitive. For a software artifact, generally only the procedural and perhaps elements of the declarative knowledge associated with a program are represented within it; capturing the other elements requires additional documentation.

## 3. INCREASING LITERATENESS OF KNOWLEDGE EMBEDDED IN SOFTWARE

In traditional programming, a number of mechanisms have been employed to capture programmer knowledge, principally various forms of documentation internal and external to the source code. Certainly the most common method is the use of comments within the code, and consequently, comments are the most relied upon source of knowledge for programmers attempting to understand previously written code [Souza et al., 2005; Das et al., 2007]. However, programmers are inconsistent in how they write comments [Marin, 2005], and efforts to structure comments have largely failed. An alternative is to provide separate (from the source code) documentation; this approach suffers from a gradual divergence from the source, as the documentation is often not maintained [Forward and Lethbridge, 2002], and as such, programmers rely on these artifacts much less than source code comments [Souza et al., 2005].

In both cases – adding structure to comments and maintaining external documentation – the challenge is a human one: how are programmers to be encouraged to maintain documentation at a level that promotes effective reuse? An approach that has emerged over the last decade is the use of automatic documentation generators, an example of which is Javadoc. This system takes comments embedded within the code and creates a hyperlinked document describing the classes and methods of a Java software artifact [Kramer, 1999]. However, as Raskin [Raskin, 2005] argues, automatically generated comments likely encourage developer laziness by giving the appearance that the documentation has been written, and he appeals to programmers to take to heart Knuth's essay on literate programming [Knuth, 1992] as the "...gospel... for all serious programmers" regarding documentation.

The literate programming approach, so named by Knuth because it engenders the perspective of software code as a form of readable literature, suggests a different approach to development, one in which the author of a program communicates the meaning of the program to another person rather than simply to a computer to execute. This is achieved by embedding segments of code within a narrative structure that describes them. Two key concepts intrinsic in this are *chunking* and *verisimilitude*. Chunking of logical code segments allows the author to describe in a natural ordering (rather than, for example, a compilation or execution ordering) the flow of the program, perhaps in the context of how it solves a problem. Chunking along logic boundaries in traditional programming has been shown to improve programmer comprehension [Norcio and Kerst, 1983]. Verisimilitude is the unification in a single artifact of the code and the documentation, which, as previously discussed, results in improved synchronization and maintenance of the documentation.

Research into the efficacy of literate programming has demonstrated positive, if occasionally anecdotal, results. Despite this promise, literate programming certainly has not become commonplace; requirements such as learning an additional markup language and a lack of programmer-friendly tools integrated into the software development cycle have been cited as impediments [Cordes and Brown, 1991]. Recent developments in Integrated Development Environments have begun to address these criticisms, and new thinking about the application of literate programming has led to modernized approaches [Pieterse et al., 2004]. Holmes [2003] offers the concept of perspicuous programming that extends the literate paradigm by explicitly incorporating documentation for a program's users within the software, that is, the software itself becomes the reference manual. When applied to the domain of end-user programming, in which the developer and user are increasingly synonymous, the resulting artifact becomes a unified one containing the software, its design rationale and the problem-solving knowledge behind the solution it encodes, and end-user literate programming is realized [Dinmore and Norcio, 2007].

**4. MODEL OF ANNOTATION IN END-USER PROGRAMMING**

As a means toward achieving end-user literate programming, this work adopts the concept of annotations, which are information provided in addition to that which is critical to the core function of an artifact; annotations here are not restricted to text, but may be "rich," containing any kind of media. Marshall [1998] identifies several dimensions for annotation including tacit to explicit and informal to formal. Tacit annotations represent incomplete thoughts, and are normally written as "memory joggers" intended to be personal and therefore useful primarily to the author. At the other end of the scale, explicit annotations contain all of the information necessary to understand the annotated thought.

With the goal of enabling increased literacy in end-user programs, a collection of design guidelines for supporting rich annotation have been formulated for the designers of end-user programming environments. As a framework for explicating these guidelines, a model of annotation in end-user programming is offered (figure 1):

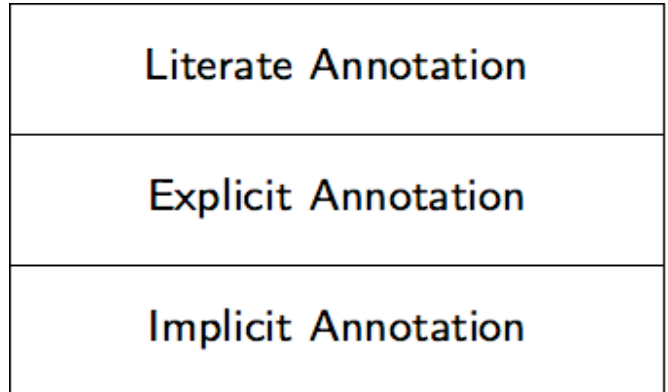

Figure 1: 3-level design model of annotation in end-user programming.

This model describes three levels of annotation: implicit, explicit and literate. The implicit level applies to annotations that are largely unwritten (as text) or only represent partial thoughts, with the rest of the author's thought remaining tacit [Marshall, 1998]. As an example in the spreadsheet paradigm, consider the programs represented by figure 2:

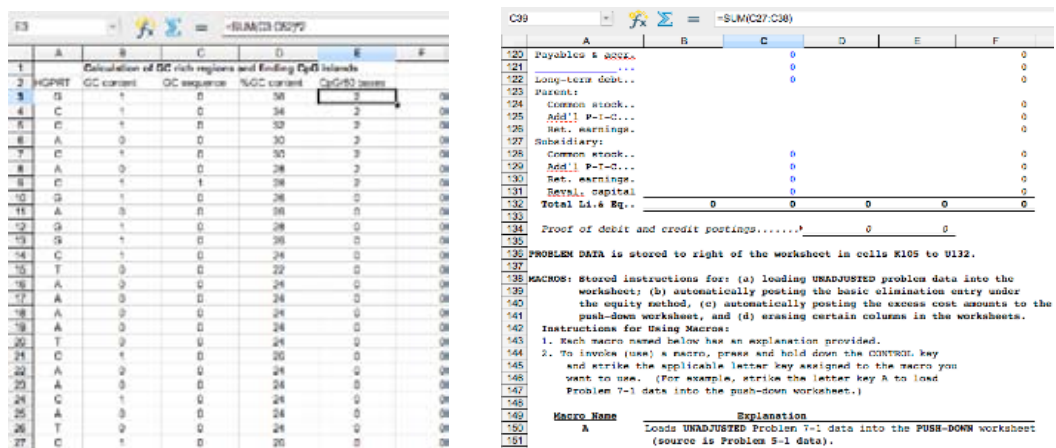

Figure 2: Implicit (left) and explicit (right) annotation in spreadsheets.

The first spreadsheet is implicitly annotated: it contains its formulae – accessible only by clicking on each cell – as well as some column headings and a sheet title. There are no instructions for using the sheet, nor are there extended descriptions of the computations.

The second sheet in figure 2 is explicitly annotated. The explicit level demonstrates intent on the part of the developer to create meaningful annotations. Many authoring environment offer mechanisms for adding text or rich, multi-media annotations that exist outside the primary medium, though the question is whether these ad hoc tools are effective. In particular, it is often left to the author to decide how to implement these to create truly communicative annotations. The results will vary. Here, the spreadsheet developer has used labels, structure, narrative text, and even color.

However, there are still some problems with this. In order to achieve this level of annotation, the developer had to enter the narrative text in the paragraphs on separate lines, *manually* wrapping the text (see the paragraphs starting at row 138). This is a constraint of the annotations being bound to the underlying structure of the spreadsheet. A possible solution is to use the commenting feature of the spreadsheet, but this again presents its own set of limitation: the annotation remains hidden until activated, it is bound to a cell rather than existing as an independent object, and when displayed, it covers a portion of the spreadsheet.

Finally, the literate level of annotation fully adopts the literate programming paradigm, and in doing so, changes the nature of the development environment to one that is composed of a linear flow of chunks of different types. Recall that the ordering of the chunks is based on the best way to describe the program rather than any notion of an execution or compilation order. This design is easily extended to include other chunk types, for example images or other rich media: multimedia has been shown to improve comprehensibility of procedural knowledge texts [Stone and Glock, 1981].

To establish the extent to which implicit and explicit annotation are present in current spreadsheet usage, a survey of a large collection of spreadsheets – the EUSES corpus [Fisher and Rothermel, 2005] – is underway. Manual sampling (n=104) preliminary to a complete survey indicates that, of those spreadsheets that are computational (41%), 42% are implicitly annotated and 58% are explicitly annotated. Of the explicitly annotated spreadsheets, about half (28% of the total) exhibited literate-like levels of annotation and chunked complexity, though they clearly were not constructed in the literate paradigm. The relatively more sophisticated use of annotation indicates a desire among users to annotate their spreadsheets, supporting the direction of this work toward developing better annotation-enabling features.

**5. PROPOSED ANNOTATION GUIDELINES**

Given this model, design guidelines for the informed-explicit and literate annotation levels can now be enumerated. These guidelines are briefly described here, and then subsequently applied in more detail to the spreadsheet paradigm.

These proposed guidelines break into two groups. The first, encompassing guidelines 1-7, are overall guidelines for the the design of annotation-supportive environments. Guidelines 8-11 suggest features that can enhance and leverage the presence of rich annotation to improve the user experience and thereby support increased adoption.

**5.1 Document problem-solving knowledge in the most accessible manner**

This guideline is the most fundamental, pragmatically calling for the use of notations that are readily accessible to the vast majority of users; in general, this means the use of textual narrative and images rather than some more compact, though less accessible notation, such as a purely mathematical or machine-centric notation. Rich text and multimedia support improved knowledge transfer [Agresti, 2000].

**5.2 Address all levels of end-user development**

M¿rch [1997] defines three levels of end-user tailoring: customization, integration and extension. Customization involves the ability to change parameters in a program (for example, values in the cells of a spreadsheet), while integration involves adding structures within the notation (adding a new spreadsheet within a workbook or new collection of formulas). Extension is about adding new features to the environment, for example through a plugin interface or macro programming language. Supporting this spectrum of users and needs is necessary for a tool to be successful.

**5.3 Seamlessly maintain documentation and software code**

This is an expression of the concept of verisimilitude [Wyk, 1990] and recognizes that unifying the software and narrative-expressed knowledge in a single, logical artifact is more effective than attempting to maintain multiple artifacts. Note that this does not require this to be done in a single logical artifact, but rather only at the presentation level. However, it is important to consider management of the artifact beyond the runtime environment; a single logical artifact is likely easier for users to manage.

**5.4 Manage annotations and code as first class objects**

Handling annotations in the same manner as the software elements places them on equal footing and presents to the user a common interaction paradigm [Marshall, 1998]. It also allows for increased interoperability among objects and a closer mapping to the users' mental model [Repenning and Ioannidou, 2006]. Borrowing from literate programming, this is accomplished by viewing all objects as chunks.

**5.5 Arrange annotations in a structured manner**

This guideline implements the finding that book-like presentations improve software comprehension and maintenance [Shum and Cook, 1994]. The linear design presents a logical flow, though in this case, the logical ordering may be in the explanation rather than the computation. This guideline is in response to explicit but uninformed designs in which the annotations have no particular unifying structure, a situation which might detract from comprehension.

**5.6 Support global visibility and ease of navigation**

Literate programs enable the automatic creation of a table of contents, cross-reference and index. Recalling the benefits of a hierarchical structure for problem-solving, and implementing the user interaction guideline of overview plus details on demand [Shneiderman, 1997], this guideline calls for the interface to include these automatically generated navigational mechanisms.

**5.7 Support multiple audiences with differing needs**

End-user developed software may serve a variety of users with different objectives ranging from neophytes who need to learn the domain, to advanced users who simply want to run a computation. These users will require different interfaces, and in the case of annotated software, different levels of annotation. This can be accomplished by facilitating themes [Kacofegitis and Churcher, 2002] and enabling chunks to be assigned to different, named themes. A user can select a theme, which then presents only the chunks in that theme. Additionally, ordering in themes can be different, allowing for different logical paths through the artifact.

**5.8 Provide scaffolds for effective problem solving and task execution**

This is the first of several higher-level guidelines, implemented on the foundation provided by the preceding guidelines. Here, it is recognized that certain structures – templates or compositions of components – can address particular use cases more effectively. For example, it may be useful to adopt a software patterns structure for certain kinds of problems. This would involve a templated arrangement of chunk types that would be tailorable, of course, but provide a scaffolding with which to start. Worked problems [Renkl et al., 2000] and task-specific designs are included within this guideline.

**5.9 Encourage documentation by automatically identifying and creating placeholders for items to annotate**

Even when well intentioned, a user is often more focused on problem-solving than documenting. By identifying key elements to annotate and creating documentation stubs for them, the user is both reminded of the need for documentation and intelligently assisted in selecting the pieces of the software artifact to document. This extends the behavior of documentation generators, such as JavaDoc [Kramer, 1999], into end-user programming.

**5.10 Prevent software errors**

Bringing software engineering methods to end users is challenging, particularly when one objective of end-user programming is to make the programming aspect essentially invisible. This guideline calls for the value-added use of the structures and rich content contained in the artifact to help prevent errors. Common features in modern programming environments, such as syntax coloring and autocompletion should be implemented. Additionally, annotations can allow for the expression of assertions as part of the documentation, and these can be actively checked at runtime [Burnett et al., 2003].

**5.11 Provide integrated, intelligent access to reusable assets**

Recommender systems can be employed to actively suggest reusable elements, thereby leveraging the rich semantics inherent in the annotations – both in the problem-solving domain and in the software domain – that may be relevant to the user's work. This also supports the development of communities; end-user programmers benefit from the artifacts produced by others and the social interaction in this process [Nardi, 1993].

**6. ANNOTATION-SUPPORTIVE SPREADSHEETS**

In this section, the above guidelines are applied to the spreadsheet modality of end-user programming. Spreadsheets are an attractive choice for an initial design-in-use evaluation: First, they are by far the most common end-user programming tool [Jones et al., 2003].Their interface semantics are relatively uniform across varying spreadsheet applications, so the guidelines will be widely applicable. Finally, they have been the subject of a significant amount of research that can be leveraged toward this objective.

The guidelines are applied here at both the informed-explicit and literate levels of the end-user programming annotation model. The key difference is that, at the explicit level, the familiar spreadsheet application is used as the foundation and features supporting annotation consistent with the guidelines are added on top of that foundation. At the literate level, a completely different application foundation (from the users' perspective) is introduced, though as argued by Sajaniemi [2002], users' familiarity with spreadsheets and text editing makes the leap much less intrusive.

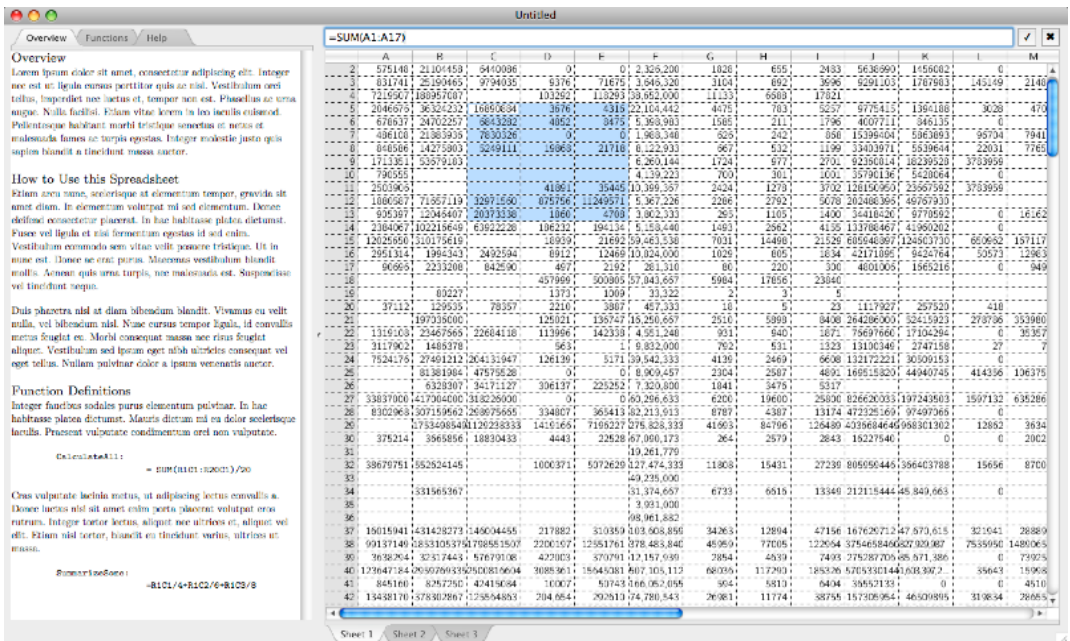

Figure 3: Informed-explicit design concept.

The informed-explicit design consists of a traditional, multi-tabbed spreadsheet stack positioned next to a multi-tabbed document pane; this pane can be closed, making the system resemble the traditional spreadsheet and potentially easing transition to the environment. The annotations pane allows for different, themed sequences of rich annotation chunks to be presented. A hyperlinked table of contents can be revealed for quickly navigating to a topic. Formulas can be defined within the text and then used in spreadsheet, providing a low barrier to rapid reuse for the user and encouraging adoption.

Extending this to the literate design, the spreadsheet pane is removed and now the spreadsheet components become another chunk type within the themed views. While a simple change to make in the design, it fundamentally changes the user's view from one in which the annotations are an easily removable – and therefore ignorable – addition to the spreadsheet, to one in which they are fully integrated.

These designs drive certain requirements for the calculation environment. For example, because there are multiple chunks, it is necessary to provide a global referencing mechanism. Properly implemented, this allows for a number of unusual features, including the ability to define formulas within a chunk-type of their own, so that the formula is entirely visible, annotatable and reusable by reference; and to reference spreadsheet ranges from within the text. The spreadsheet execution paradigm is an excellent match with the literate design, because it allows for the chunks to be presented in any order. Since calculations only rely on the reference graph, the presentation can be completely decoupled from the computation.

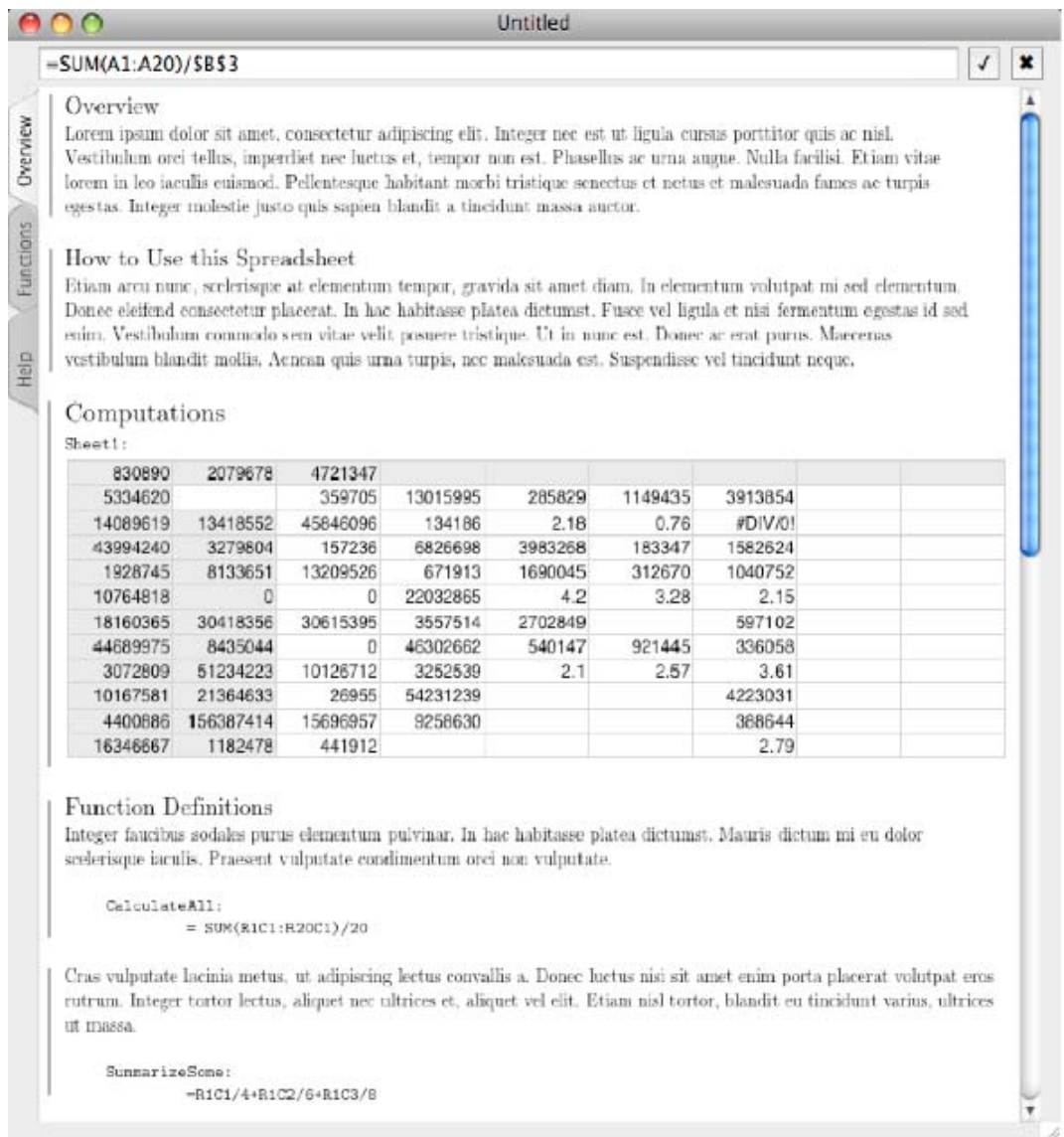

Figure 4: Literate design concept.

**7. EVALUATION**

Studies are in progress to evaluate designs based on the proposed guidelines as compared to traditional spreadsheets; the initial research is designed to explore the efficacy of the proposed guidelines in improving knowledge transfer and reuse. To accomplish this, the Location-Comprehension-Modification model of reuse [Fischer et al., 1991] provides a framework for an experimental study in which subjects are measured in each of the aspects of this model across four conditions: implicit annotation (traditional spreadsheet with no annotations), uninformed-explicit (traditional spreadsheet with use of built-in commenting tools and in-cell annotations), informed-explicit (a new, guideline-informed design based on the spreadsheet paradigm), and the new, literate design. Subjects will perform three experiments: the first will measure the relative findability [Azzopardi, 2008] of artifacts in a corpus ("location", where the hypothesis is that rich annotation increased intrinsic findability using common search engines). The second will measure comprehension using classical techniques, such as simulation [Dunsmore and Roper, 2000]. Finally, subjects will perform a maintenance task to measure their ability to modify an artifact. It is expected that the individual and aggregate results from each sub-experiment will be measurably different among conditions, with the design interventions demonstrating better results. Should these results support the claim of increased reuse, future studies will focus on optimizing the interaction mechanisms for creating rich annotations, thereby further refining the guidelines and offering specific, experience-based implementation tradeoffs to software designers.

Throughout the initial work, and certainly in the annotation creation interaction-specific work, the Cognitive Dimensions of Notations framework [Green, 1989] will be used throughout to guide the designs and their evaluation from a notational usability perspective. This framework is appropriate in that it has been demonstrated to effectively address end-user programming and, more generally, the abstract idea of notations.

## 8. RELATED WORK

There are a number of previous research efforts focused on improving the documentation of spreadsheets. Several authors approach the challenge in the context of existing spreadsheet tools, primarily Microsoft Excel, offering recommendations for structures and best practices in support of documentation. Payette [2008] describes a methodology for applying the built-in "comment" functionality to create structured annotations. Powell and Baker [2007] present a chapter on spreadsheet engineering best practices, among them "designing for communication" and "documenting important data and formulas." Similarly, Raffensperger [2008] references the structure of traditional programs in introducing a similarly-inspired structure for spreadsheets, summarized in four general guidelines, including "format for description, not decoration" and "expose rather than hide information."

Sajaniemi [2002] describes a prototype implementation of a fusion of the the traditional spreadsheet and word processor. This design presents a unified user interface that naturally combines many of the features to which users are already accustomed. Among the functionalities that would need to be reconsidered for this combined paradigm are the display of results, the indexing notation, and referencing mechanism.

Paine [2008] offers a proposal to extend the Excelsior [Paine et al., 2006] spreadsheet design language with literate programming-style documentation constructs. Excelsior is inherently modularized, and as such, fits the chunked style of literate programming well. This approach is more in keeping with the original processing model for literate programming, as well: programs written in Excelsior exist as separate text files that are then used to generate a spreadsheet and, in the case of Literate Excelsior, documentation. In the present work, there is no structurally separate processing phase.

## 9. CONCLUSION

This work seeks to improve the capture and transfer of problem-solving knowledge in end-user programming. Efforts to date have focused on the explication of a collection of design guidelines for seamlessly incorporating explicit annotation into various end-user programming modalities, with an initial focus on the widely-used spreadsheet. These design guidelines draw heavily from the literatures of human-computer interaction, software engineering, knowledge management and design of communications, as well as pragmatic studies of the spreadsheet paradigm, and are strongly influenced by Knuth's literate programming. Design prototypes are in the process of being iteratively improved and an experimental evaluation of the designs with respect to their impact on artifact reusability and user satisfaction is planned.